\documentclass[12pt,a4paper]{article}
\usepackage{amscd,amsmath,amssymb,bbm}
\usepackage[english]{babel}
\usepackage{graphicx}

\numberwithin{equation}{section}

\renewcommand{\L}{\Lambda}
\newcommand{\la}{\lambda}
\renewcommand{\t}{\tau}
\newcommand{\G}{\Gamma}
\renewcommand{\a}{\alpha}

\renewcommand{\o}{\omega}
\renewcommand{\th}{\theta}

\newcommand{\Z}{\mathbbm{Z}}
\newcommand{\R}{\mathbbm{R}}
\renewcommand{\P}{\mathbbm{P}}
\renewcommand{\H}{\mathbbm{H}}
\newcommand{\C}{\mathbbm{C}}

\renewcommand{\l}{\left(}
\renewcommand{\r}{\right)}
\newcommand{\dd}{{\mathrm d}}
\newcommand{\diff}[1]{\frac{{\mathrm d}}{{\mathrm d}#1}}
\newcommand{\ddiff}[1]{\frac{{\mathrm d}^2}{{\mathrm d}#1^2}}

\newcommand{\bm}{\boldsymbol}

\newcommand{\2}{\frac{1}{2}}                          
\newcommand{\4}{\frac{1}{4}}
\newcommand{\mc}[1]{\mathcal{#1}}
\newcommand{\be}{\begin{equation}}
\newcommand{\ee}{\end{equation}}
\newcommand{\<}{\langle}
\renewcommand{\>}{\rangle}

\renewcommand{\i}{\mathrm i}

\begin{document}


\begin{titlepage}
\setcounter{page}{0}
\begin{flushright}
      hep-th/0607142\\
      ITP--UH-15/06\\
\end{flushright}
\vskip 2.0cm

\begin{center}
{\LARGE\bf  Pure Gauge $\bm{SU(2)}$ Seiberg-Witten Theory and Modular Forms}
\\
\vspace{14mm}

{\Large
Kirsten Vogeler}
\ \ \ and \ \ \
{\Large
Michael Flohr}
\\[5mm]
{ \em
Institut f\"ur Theoretische Physik, Universit\"at Hannover \\
Appelstra\ss{}e 2, 30167 Hannover, Germany }
\\[5mm]
email: \texttt{vogeler, flohr @itp.uni-hannover.de}
\\[12mm]
\small \today
\end{center}
\vspace{15mm}


\begin{abstract}
\noindent
We identify the spectral curve of pure gauge $SU(2)$ Seiberg-Witten theory
with the Weierstrass curve $\C/L\ni z\mapsto (1,\wp(z),\wp(z)')$ and thereby
obtain explicitely a modular form from which 
the moduli space parameter $u$ and lattice parameters $a$, $a_D$ can be 
derived in terms of modular respectively theta functions. We further discuss 
its relationship with the $c=-2$ triplet model conformal field theory.
\end{abstract}

\vfill
\end{titlepage}



\pagebreak


\section{Introduction}
The low energy effective $\mc{N}=2$ SYM theory is mathematically and 
physically extremely rich and a crosspoint for string theory, topological 
field theory, Riemann surfaces, algebraic topology and other interesting 
topics \cite{Klemm:1997gg, Lerche:1996xu, Seiberg:1994rs, Seiberg:1994aj}. 
\par
In this paper we draw our attention mainly to the correspondence between
pure gauge $SU(2)$ SW-theory and the Weierstrass formulation of elliptic
curves. In particular we identify the spectral curve 
with the Weierstrass curve and thereby
obtain some means for writing the parameter of the moduli space and the
lattice parameters $a$ and $a_D$ in terms of theta functions, as was
originally proposed in \cite{Nahm:1996di}. The basic motivation for doing this
is that theta functions may point to characteristic functions which have a 
meaning
as amplitudes in conformal field theory. Indeed, we find that the modular
form from which we derive $a$ and $a_D$ can be expressed by
characteristic functions of the $c=-2$ triplet model. The latter is a nice 
tool in order to describe the 
Weierstrass curve and its developing parameter.
\par
The organization of the paper is as follows. First we review how the torus,
the spectral curve for pure gauge $SU(2)$ SW-theory, is related to the
charge lattice. Then we identify the corresponding algebraic curve with 
the Weierstrass elliptic curve. In the last chapters we explain 
how we derived the modular form mentioned above and then discuss the 
relation to the $c=-2$ triplet model. A conncetion of SW-theory to
loagrithmic conformal field theory has often been speculated, see for example
\cite{Cappelli:1997qf,Flo98,Flohr:2004ug}. 
We argue that indeed the modular forms in SW-theory
can only be expressed in terms of characters or torus amplitudes of a conformal
field theory, if this theory is logarithmic.


\section{From Pure Gauge $\bm{SU(2)}$ SW-Theory to the Torus}
In pure gauge SW-theory, the spectrum of stable, charged particles is given 
by points on a scale dependent lattice spanned by the vacuum expectation
values $a(u)$ of 
the scalar field $\phi$ and its dual $a_D(u)$: $L=(a_D,~a)^T$. The parameter 
$u$ denotes a point in the moduli space 
$\mc{M}:=\C\P^1\setminus\{\L^2,-\L^2,\infty\}$ of 
the theory, where $\L$ usually gets related to the scale $\L_{QCD}$,
$\L\in\R^+$ and 
$L:\mc{M}\rightarrow E$ can be interpreted as a section of some flat 
holomorphic $\G$-bundle $E$, $\G\subset SL(2,\Z)$. The 
motivation to 
consider such a lattice comes from the BPS- respectively central charge formula
\be\label{cc}
  Z=n_e~a+n_m~a_D,
\ee
where $n_e$ and $n_m$ are the quantum numbers for the electric and magnetic
charges. $\G\subset SL(2,\Z)$ is the subgroup, acting on $L$, which 
leaves $Z$ invariant. 
It is determined by the local symmetries of the theory which are due to the
possible monodromies of $L$ around the singularities of $\mc{M}$ and the
global $\Z_2$ symmetry $u\mapsto -u$ on $\mc{M}$.
\par
Geometrically the charge lattice has an interpretation as the cycles of
the spectral curve of SW-theory. In particular
there exists a mapping $\la_{SW}$ of $\mc{M}$ to the space of meromorphic 
one-forms, called the SW-form, encoding the information of $L$ and 
satisfying 
\begin{align}\label{aad}
  \begin{split}
    &\la_{SW}(u):~L(u)\rightarrow~H_1(\mc{E}(u),\mathbbm{Z}),\\
    &(a_D(u),~a(u))=(\oint_B\la_{SW}(u),~\oint_A\la_{SW}(u))~~~
    A,B\in~H_1(\mc{E}(u),\Z).
  \end{split}
\end{align}
One now can obtain the spectral curve as a Jacobian variety from the periods
\be
  (\Pi_D(u),~\Pi(u))=(\diff{u} a_D,~\diff{u} a)=
  (\oint_B\o(u),~\oint_A \o(u))~~~ A,B\in~H_1(\mc{E}(u),\Z).
\ee
Hence, to every point $u\in\mc{M}$ we attach a torus $\mc{E}(u)$ with
periods $(\Pi_D(u),~\Pi(u))$ and developing mapping
\be\label{taudef}
\t_{SW}(u)=\frac{\Pi_D(u)}{\Pi(u)}=\frac{\dd a_D(u)}{\dd a(u)},
\ee
which is an element of the complex upper half plane $\H$ and coincides with 
the coupling constant of SW-theory. Since the local physical information is 
preserved under the symmetry $\G$, which has an action on $L$ and on 
$\t_{SW}$ via
(\ref{taudef}), the moduli space of the theory in the perspective of the 
torus thus can be obtained as the quotient of the uniformization space
$\H$ with $\G$:
\be
  \tilde{\mc{M}}=\H/\G.
\ee
The relations as described above can be summarized as follows: 
\be
\begin{array}{*3{c}}
  L&\stackrel{\la_{SW}}{\rightarrow}&H_1(\mc{E},\mathbbm{Z})\\
  \downarrow&&\downarrow\\
  \mc{M}&\stackrel{\t_{SW}}{\rightarrow}&\tilde{\mc{M}}\\
\end{array}
\ee
$\t_{SW}$ is a bi-holomorphic mapping and the singularities of 
$\mc{M}$ correspond to the vertices in $\tilde{\mc{M}}$.
The action of $\G$ on $H_1(\mc{E},\Z)$ is promoted 
by $\la_{SW}$ and can be interpreted as a change of basis which of course is a 
mapping between equivalent tori. 


\section{The spectral curve in terms of {\boldmath $\t_{SW}$}}
In this section we will relate the torus $\mc{E}(u)$, i.e.\ the spectral
curve of $SU(2)$ SW theory to the Weierstrass standard form and thus change
our perspective from the moduli space $\mc{M}$, parametrized by $u$, to the 
moduli space $\tilde{\mc{M}}$, parametrized by $\t_{SW}$ or from holomorphic
sections to modular forms. This has been done in several publications
such as \cite{Nahm:1996di}, \cite{Huang:2006si} and \cite{Aganagic:2006wq}.
However, the aim of this section is twofold, on the one hand we define all 
quantities we have used for the succeeding sections and on the other hand
we focus on the relation of the uniformization parameter $\t$ to 
$\t_{SW}$. The latter is important as it makes a first connection of
the $c=-2$ triplet model with $SU(2)$ SW theory, that has already been observed
in \cite{Flohr:2001zs}, even more explicit.

A torus $\mc{E}(u)$ has a description as an algebraic curve\footnote{There
exists another, physically equivalent formulation of the torus related to the 
one above by an isogeny \cite{Seiberg:1994aj}, \cite{Klemm:1995wp}.}
\be\label{algcurves}
 \mc{E}(u):~~~y^2 =(x^2-u)^2-\L^4,
\ee
\noindent where we use the charge lattice normalization 
$L(u)=(a_D(u),~\2 a(u))^T$ corresponding to the coupling $2\t_{SW}$ and the 
monodromy group $\G=\G^{0}(4)$ (c.f.\ Appendix \ref{modu}).
Uniformization relates $\mc{E}(u)$ to the 
family of standard elliptic curves $\mc{E}(\t)$, described via the Weierstrass 
function
$\mathbbm{C}/L~\ni~z\mapsto~(1,\wp(z),\wp'(z))~\in~\C\P^1$ and 
parametrized by some developing parameter $\t$ in the complex upper half plane 
$\H$~:
\begin{equation}\label{weierstr}
  \begin{aligned}
  \mc{E}(\t):~~~\l\diff{z}\wp(z)\r^2
  &=\phantom{:}4\wp(z)^3-g_2(\t)\wp(z)-g_3(\t)\\
  &=:4\Pi_{i=1}^3(\wp(z)-e_i),~~~z\in H,~\Im(\t)>0.
  \end{aligned}
\end{equation}
The curve (\ref{algcurves}) is identified with (\ref{weierstr})
by first converting it to the Weierstrass standard form and then
matching $J(u)=J(\t)$, where (c.f. Appendix \ref{modu})
\be\label{J}
  J(\t):=\frac{4}{27}\frac{(\la^2(\t)-\la(\t)+1)^3}{\la^2(\t)(\la(\t)-1)^2}.
\ee
Here we use a general parameter $\t$ that will be related to $\t_{SW}$, soon.
For one of the positive roots of $u^2$ this identification yields:
\begin{equation}\label{us}
     u(\t)=\frac{\L^2}{2}~\frac{\la(\t)+1}{\sqrt{\la(\t)}}.
\end{equation}
It is possible to invert this equation in order to derive an expression 
for $\la$ in terms of $u$ but another equivalent and physically inspired way 
is to identify $\la$ with the inverse crossing ratio
of the branch points $\bar{e}_1:=\sqrt{u-\L^2},~\bar{e}_2:=-\sqrt{u+\L^2},~
\bar{e}_3:=-\sqrt{u-\L^2}$ and $\bar{e}_4:=\sqrt{u+\L^2}$ of the curve
$\mc{E}(u)$ 
\be\label{myxi}
  \la=\xi:=\frac{(\bar{e}_1-\bar{e}_4)(\bar{e}_3-\bar{e}_2)}
	 {(\bar{e}_2-\bar{e}_1)(\bar{e}_4-\bar{e}_3)}
\ee
in the $c=-2$ triplet model on the sphere, as it is done 
in \cite{Flohr:2004ug}. This is possible due to relation (\ref{tauinhyp})
which canonically appears in the $c=-2$ triplet model as a quotient of
certain correlation function (c.f.\ (\ref{mucorr})). We will comment on that 
more extensively in chapter $5$. Notice that thereby we relate the triplet
model to the Weierstrass formulation but if $\t_{SW}=\t$ this also means a
connection to SW theory. Simplifying (\ref{myxi}) yields
\begin{equation}
       \la(\a) =
       \frac{\sqrt{\frac{\a}{\a-1}}-1}{\sqrt{\frac{\a}{\a-1}}+1}~~, 
       ~~\a=\frac{u^2}{\L^4}.
\end{equation}
In order to identify $\t$ and $\t_{SW}$ use the fact that by means of 
uniformization the structure of the algebraic curve (\ref{algcurves}) is 
encoded in a
hypergeometric differential equation for the charge lattice parameters
$a$ and $a_D$ (f.i.\ \cite{Bilal:1995hc}, 
\cite{Klemm:1997gg}, \cite{Lerche:1996xu}):
\be
  \hat{\mc{L}}~s(u):=\l(\L^4-u^2)\ddiff{u}-\4\r~s(u)=0.
\ee
The two solutions are \cite{Bilal:1995hc}
\begin{equation}\label{dgla}
  \begin{aligned}
    a(u)&=\sqrt{\frac{u-\L^2}{2}}~{}_2F_1\l\2,-\2;1;\frac{2\L^2}{\L^2-u}\r,\\
    a_D(u)&=-\frac{\L^2+u}{2\L}~{}_2F_1\l\2,\2;2;\frac{\L^2+u}{2\L^2}\r.
  \end{aligned}
\end{equation}
Now using (\ref{taudef}) and (\ref{tauinhyp}) yields $\t=\t_{SW}$.

Expressions of the scalar modes in terms of hypergeometric functions are,
of course, not new. However, we provide these here in order to make the
paper more self contained. 


\section{Modular Forms}
In order to get in contact with cfts or string theory it is suggestive to 
express the
main functions above as modular forms (c.f.\ Appendix \ref{modu}). To be
more specific, in our case this is important as some of the characters of the 
triplet model have a characteristic inhomogeneity with respect to their 
modular weight. We will find that the main character of this section
(\ref{c_tau}) can be expressed in terms of characters of a cft that has to 
have the same properties and hence is the $c=-2$ triplet theory.

The moduli space parameter $u$ 
can be written in terms of the Dedekind eta function from (\ref{us}):
\be
   u(\t)=\frac{\L^2}{8}~ \l \l\frac{\eta(\frac{\t}{4})}{\eta(\t)}\r^8+8\r.
\ee
This is the key for deriving all other quantities as functions of $\t$. 
In \cite{Nahm:1996di} Nahm pointed out that the combination 
$c(\t):=\t a-a_D$ transforms 
like a modular form of weight $-1$. Another promising feature of this 
combination is that given $\t_{SW}=\t$ one yields $a$ by differentiating
$c$ by $\t$
\be
  \frac{\dd c}{\dd\t}=a+\frac{\dd a_D}{\dd\t}\l \t-\frac{\dd a_D}{\dd a}\r=a.
\ee
Hence $a$ and $a_D$ should be given in terms of modular functions, likewise.
\par
By inserting $u(\t)$ into (\ref{dgla}), $a$ and $a_D$ are expressed as
functions of $\t=\t_{SW}$. In order to write them as modular
functions it suffices to consider $a$ and $a_D$ in some region of
$\mc{M}$, they can then be analytically continued to the whole of $\mc{M}$. 
Hence, we expand $a$ and $a_D$ around $u=\infty$ and then insert
$u(\t)$ in order to find the $q$ expansions. The $u$ expansion of $a$ at 
infinity is straightforward whereas for $a_D$ one has to use \cite{Bateman} 
\begin{equation}
  \begin{aligned}
  \frac{\G(a)}{\G(c)}~{}_2F_1(a,a;c;z)&=
  \frac{(-z)^{-a}}{\G(c-a)}~\sum_{n=0}^{\infty}\frac{(a)_n(1-c+a)_n}
  {n!^2}~z^{-n}\l \log(-z)+h_n\r,\\
  h_n&=2\psi(1+n)-\psi(a+n)-\psi(c-a-n).
  \end{aligned}
\end{equation}
Using inspiration from Nahm's paper and maple we were able to match 
(analytically) the $q$ expansion of $a$ with the $q$ expansion of the 
derivative of some function $c$ by $\t$, where
\be\label{c_tau}
  c(\t):=\frac{\i\L}{\pi}
  \l\frac{\theta_{02}(\t)-\theta_{22}(\t)}{\eta^{3}(\t)}\r
\ee
in accordance with Nahm and, indeed,
\be
  a(\t)=\diff{\t}c(\t).
\ee 
What relates expression (\ref{c_tau}) and, it being the central element 
generating $a$ and $a_D$, the whole story to the $c=-2$ 
triplet model is its modular weight. The characters of ordinary 
rational cft's are homogeneous functions of modular weight zero, however
$c(\t)$ has weight $-1$. Thus, using $\chi_{-\frac{1}{8}}$ and 
$\chi_{\frac{3}{8}}$ (c.f.\ Appendix \ref{modu}) one finds
\begin{equation}
 c(\t)=\frac{\i\L}{\pi}\l\frac{\chi_{-\frac{1}{8}}-
   \chi_{\frac{3}{8}}}{\eta^2}\r
\end{equation}
and the interesting factor of $\eta^2$ in the denominator.
This, however, can only be expressed in 
terms of characters of the $c=-2$ triplet model as only there characters 
appear with inhomogeneous modular weight, and which are of the form 
$\frac{\theta_{s.t.}}{\eta}+k\eta^2$ (c.f. 
Appendix \ref{modu}).

When expressing $a_D$ by means of $c$ and $a$ we found a relation 
which differed from Nahm's given above, namely
\be
  a_D(\t)+2a(\t)=\t a(\t)-c(\t).
\ee
Anyway, we can use the $\Z_2$ symmetry $u\mapsto -u$ as it has an effect on 
$a_D$. Indeed, when going to $-u$ we calculated in exactly the same 
way as above
\be\label{anormal}
  a_D(\t)=\t a(\t)-c(\t),
\ee
and obviously
\be
  u\mapsto -u~~~\Rightarrow~~~ a_D\mapsto a_D-2a,
\ee
in accordance with, for example, \cite{Lerche:1996xu}. Further, with these 
relations and also with the help of the series expansions obtained we
succeeded in checking, again
\be
  \t_{SW}=\frac{\dd a_D}{\dd a}=\t.
\ee


\section{Correspondence to the {\boldmath $c=-2$} Triplet Model}

As we discovered in the section before, equation (\ref{c_tau}) gives a
strong hint that SW theory is related to the $c=-2$ triplet theory. We will 
now give some more relations concerning the main quantities of pure gauge
$SU(2)$ SW theory. Thereby all important parameters are obtained as
amplitudes and characters of the triplet model.

First of all, it is possible to write $c(\t)$ completely in terms of 
characters of the $c=-2$ triplet theory (on $\C\P^1$): 
\be\label{cchar}
  c(\t)=\frac{\i\L}{\pi}\l\frac{\chi_{-\frac{1}{8}}(\t)-\chi_{\frac{3}{8}}(\t)}
  {\chi_1(\t)-\chi_0(\t)}\r.
\ee
We consider this very simple expression for $c(\t)$ in terms of the
characters of the triplet theory as the main result of the paper.
We once more stress that it is not possible to express the modular form
$c(\t)$ of weight $-1$ in terms of characters or torus amplitudes of any
ordinary (rational) conformal field theory as these necessarily have to
be modular forms of weight zero. Only within logarithmic conformal field
theory do torus amplitudes arise, which have non-vanishing modular 
weight, see \cite{Flohr:2005cm}. 

There is some geometric interpretation of this behind it which is strongly 
suggested in \cite{Flohr:2004ug} and \cite{Knizhnik:1987xp}. The involved 
characters in the numerator belong to the twistfield $\mu$ of conformal 
weight $h_{\mu}=-\frac{1}{8}$ 
and its excited partner $\sigma$ with weight $h_{\sigma}=\frac{3}{8}$. 
Now, the $\mu$ 
field can be thought of as representing some branch point on $\C\P^1$ and one
may think of the torus as being constructed as a double cover of $\C\P^1$ with
four fields $\mu$ producing two branch cuts. 
\par
However, as we learn from
\cite{Knizhnik:1987xp}, this geometric interpretation of branch points being 
represented by some analytic fields is not a speciality of the triplet theory 
alone but of a whole class of theories of pairs of analytic anticommuting 
fields of integer spins $j$ and $1-j$. Nonetheless the triplet theory is 
somewhat outstanding. All these theories lead to different kinds of Coulomb gas
models according to which one can calculate different four-point functions.
The main point is that the four-point function for the
``geometric'' field $\mu$ in the $c=-2$ triplet theory leads to a
hypergeometric ode of type\footnote{c.f. \cite{Gaberdiel:2001tr}.} $(\2,\2,1)$
and hence to the projective solution (\ref{tauinhyp}), formally 
\be\label{mucorr}
  \t=\i~\frac{\<\mu(\infty)\mu(1)\mu(\xi)\mu(0)\>_2}
             {\<\mu(\infty)\mu(1)\mu(\xi)\mu(0)\>_1},
\ee
where the index labels the different conformal blocks respectively homology
cycles, that are, as explained, in one to one correspondence to the periods,
and the crossing ratio $\xi$ matches the modular function $\la$, as in
(\ref{myxi}). 
Explicitly, they read $\<\mu\mu\mu\mu\>_k = [\xi(1-\xi)]^{1/4}F_k(\xi)$
with $F_1(\xi) = {}_2F_1(\frac{1}{2},\frac{1}{2};1;\xi)$ and $F_2(\xi)=
F_1(1-\xi)$.
In this respect the
triplet theory is a natural tool in order to express the family of 
Weierstrass curves $\mc{E}(\t)$ by means of four point functions of the
twistfield $\mu$. 
\par
The relation to SW theory is then introduced indirectly via the isomorphism
$\mc{E}(u)\simeq\mc{E}(\t)$ under which $\t_{SW}=\t$ and the $c=-2$ model is
thus natural for the Weierstrass formulation with developing mapping
(\ref{tauinhyp}) rather than for the SW moduli space in general. This is
mainly due to the fact that the $c=-2$ theory does not know about the 
restrictions from the monodromy incorporated in the SW model, however it
may serve as a nice tool in order to describe the geometric properties of
general tori in the Weierstrass formulation. For example one can express the 
SW-form
\be\label{differ}
  \la_{{\rm SW}}=\frac{\i}{\sqrt{2}\pi}\frac{x^2}{y(x)}\dd x
\ee
and the derived lattice paramters $a$ and $a_D$ in terms of quotients of 
conformal blocks of the $c=-2$ triplet model \cite{Flohr:2004ug}, where 
one uses the algebraic curve $y(x)=\prod_{i=1}^{4}~(x-\bar{e}_i)$ defined by 
the four branch points $\bar{e}_i$ as in (\ref{myxi}). In particular
(\ref{aad}) has an expression as six point functions on the sphere 
\cite{Flo98} or in another interpretation two point functions on the torus, 
represented by the twistfields: 
\be
\begin{aligned}
 a_i(u) &= \frac{\<V_2(\infty)
                  \mu(\bar{e}_1)\mu(\bar{e}_2)\mu(\bar{e}_3)\mu(\bar{e}_4)   
		  V_{-2}(0)\>_i}
               {\<V_1(\infty)
                  \mu(\bar{e}_1)\mu(\bar{e}_2)\mu(\bar{e}_3)\mu(\bar{e}_4)   
		  V_{-2}(0)\>}\\
	 &= \frac{(\mathrm{i})^i}{\sqrt{2}\pi}
	   \frac{\bar{e}_3^2}{\sqrt{(\bar{e}_4-\bar{e}_3)(\bar{e}_2-\bar{e}_1)}}
           \frac{\<V_2(\varpi)
                  \mu(\infty)\mu(1)\mu(\xi)\mu(0)   
		  V_{-2}(\eta)\>_i}
               {\<V_1(\varpi)
                  \mu(\infty)\mu(1)\mu(\xi)\mu(0)   
		  V_{-2}(\eta)\>},
\end{aligned}
\ee
that is
\be
\begin{aligned}
   a(u)   &= \mathrm{i}\frac{\Lambda^2(1+\varpi^2)^2}{\sqrt{\varpi}
           \left(\sqrt{(\varpi+1)^2\Lambda^2}+
              \sqrt{(\varpi-1)^2\Lambda^2}\right)}
           F_D^{(3)}(\textstyle{\frac{1}{2},\{\frac{1}{2},2,-2\},1;}\varpi^2,
	   -\varpi,\varpi),\\
  a_D(u) &= \frac{\Lambda^2(1+\varpi^2)^2}{\sqrt{\varpi}
           \left(\sqrt{(\varpi+1)^2\Lambda^2}+
              \sqrt{(\varpi-1)^2\Lambda^2}\right)}
  F_D^{(3)}(\textstyle{\frac{1}{2},\{\frac{1}{2},2,-2\},1;}1-\varpi^2,
	   1+\varpi,1-\varpi).\\
\end{aligned}
\ee
The index $i=1,2$ labels the standard homology basis, i.e.\ $a$ and $a_D$, 
respectively, and the fields $V_2(\infty)$ and $V_{-2}(0)$ correspond to the 
double pole at infinity and the double zero at the origin of the 
SW-differential (\ref{differ}). The denominator serves in order to extract the 
nontrivial part of the numerator \cite{Flohr:2004ug}. That $a$ can be
written as a Lauricella function and that the three crossing ratios are 
related to each other was obtained in \cite{Akerblom}. Again, $\xi$ is 
related to $\la$ via (\ref{myxi}) and $\varpi^2=\xi$, $\eta=-\varpi$, such 
that the periods are functions of one complex variable only.
\par
By means of $u(\t)$, i.e.\ (\ref{us}), the expression for $a(u)$ as above
can be expanded in $q$ via the elliptic modulus
$\lambda(\tau)=\varpi^2$. This $q$-series coincides with the one for
the simple hypergeometric
functions as given in (\ref{dgla}). Using this form of $a(u)$, we find
\be\label{newa}
\begin{aligned}
  a(u) &= \L \l \frac{(\varpi-1)^2}{4\varpi}\r^{\2}{}_2F_1\l\frac{1}{2},
  -\frac{1}{2};1;\frac{-4\varpi}{(1-\varpi)^2}\r\\
  &= \Lambda(-\xi')^{-1/2}{}_2F_1\l\frac{1}{2},-\frac{1}{2};1;\xi'\r\\
  &= \frac{\mathrm{i}\Lambda}{\sqrt{2}\pi}
          \frac{\<\sigma(\infty)\sigma(1)\mu(\xi')\mu(0)\>_1}
          {\<\sigma(\infty)\sigma(1)\mu(\xi')\sigma(0)\>}\,,\\
  a_D(u)&=-\l 1-\frac{1}{\xi'}\r{}_2F_1\l \2,\2;2;1-\frac{1}{\xi'}\r\\
  &= \frac{\mathrm{i}\Lambda}{\sqrt{2}\pi}
          \frac{\<\sigma(\infty)\sigma(1)\mu(\xi')\mu(0)\>_2}
          {\<\sigma(\infty)\sigma(1)\mu(\xi')\sigma(0)\>}\,,\\
  &= \frac{-\Lambda}{\sqrt{2}\pi}
          \frac{\<\sigma(\infty)\sigma(1-1/\xi')\mu(1)\mu(0)\>_1}
          {\<\sigma(\infty)\sigma(1-1/\xi')\mu(1)\sigma(0)\>}\,.
\end{aligned}
\ee
Here, $\xi'$ is interpreted as the inverse crossing ratio
\be
  \xi':=\frac{-4\varpi}{(\varpi-1)^2}.
\ee 
Again, the label
of the correlation functions refers to the conformal blocks.
In the denominator, $\<\sigma(\infty)\sigma(1)\mu(\xi')\sigma(0)\> = 
[\xi'(1-\xi')]^{-1/4}(C_1\xi'+C_2)$, and the block needed here is obtained 
with the choice $C_1=1, C_2=0$. The final expression for $a_D(u)$ is obtained
by analytic continuation of the second conformal block to the region
$\{\xi':|\xi'|>1, |1-\xi'|<1\}$.
\par
Performing again an expansion in $q$ for $\xi'$,
we find explicitly the relation
\be
  \xi=\varpi^2=\lambda(q)\ \ \ \textrm{and}\ \ \ 
  \xi'=\lambda(-\sqrt{q})\,.
\ee
This means that the modular parameter associated to $\xi'$ is just half
of the modular parameter associated to $\xi$, since $-\sqrt{q}=-(\exp(\pi
\mathrm{i}\tau))^{1/2}=\exp(\pi\mathrm{i}(\tau/2\pm 1))$, and
$\tau/2\pm 1$ is equivalent under the $PSL(2,\mathbb{Z})$ action to $\tau/2$.
We note that substituting $u\mapsto-u$ would have led us immediately to
the result $\xi'=\lambda(\sqrt{q})$, which is the elliptic modular form for
a torus with lattice parameter $\2\t$. In this respect the
correlation functions in (\ref{newa}) have an interpretation as zero-point 
functions on a torus with $\2\t$, where the branchpoints are represented not 
by the $\mu$ fields alone but also by the excited fields $\sigma$.
\par
To conclude the discussion, we remark that the interpretation of the 
denominator of equation (\ref{cchar}) is 
much more involved. That $c(\t)$ can be expressed solely in terms of 
characteristic functions of the triplet model can be read as a strong hint 
that there really is a relation, though up to know we obtained all quantities
in a non-constructive way. It would be interesting to look at SW with
matter and other gauge groups in order to find out if still the triplet model 
appears. For completeness, we give the SW periods $a$ and $a_D$ expressed
in terms of characters of the $c=-2$ triplet model:
\be
\begin{aligned}
  a(\tau) &= \Lambda
  \frac{\theta_3(\theta_2^4-\theta_3^4)\partial_{q}
    \theta_2}
  {\theta_2^2(\theta_2\partial_{q}\theta_3-\theta_3
    \partial_{q}\theta_2)}\\
          &= \Lambda
  \frac{\eta^2(\chi_{-\frac{1}{8}}+\chi_{\frac{3}{8}})
  \left(16(\chi_0+\chi_1)^4 - (\chi_{-\frac{1}{8}}+\chi_{\frac{3}{8}})^4
  \right)\partial_{q}(\eta(\chi_0+\chi_1))}
  {4(\chi_0+\chi_1)^2\left( (\chi_0+\chi_1)\partial_{q}
  (\eta(\chi_{-\frac{1}{8}}
  +\chi_{\frac{3}{8}})) - (\chi_{-\frac{1}{8}}+\chi_{\frac{3}{8}})
  \partial_{q}(\eta(\chi_0+\chi_1))\right)}\\
          &= \Lambda
  \frac{(\chi_0-\chi_1)(\chi_{-\frac{1}{8}}+\chi_{\frac{3}{8}})^2
  \left(16(\chi_0+\chi_1)^4 - (\chi_{-\frac{1}{8}}+\chi_{\frac{3}{8}})^4
  \right)}
  {4(\chi_0+\chi_1)^2}\\
  &\phantom{=} \times\frac{
  \partial_{q}((\chi_0-\chi_1)(\chi_0+\chi_1)^2)}{
  \left( (\chi_0+\chi_1)^2\partial_{q}( 
  (\chi_0-\chi_1)(\chi_{-\frac{1}{8}}
  +\chi_{\frac{3}{8}})^2) - (\chi_{-\frac{1}{8}}+\chi_{\frac{3}{8}})^2
  \partial_{q}( (\chi_0-\chi_1)(\chi_0+\chi_1)^2)\right)}
  \,,\\
  a_D(\tau) &= (\tau-2)a(\tau) - c(\tau)\,,
\end{aligned}
\ee
where $c(\tau)$ is given in eq.~(\ref{cchar}) and the theta as well as the
characteristic functions
are all taken at $\sqrt{q}$, i.e.\ at the value $\2\t$. Note that this 
corresponds to $\xi'$ above, i.e.\
half of the modular parameter associated to the original formulation in $\xi$.
To arrive at 
these formul\ae, one essentially uses the facts that, firstly, $a(u)$ is
given by 
\be
a(k) =\frac{2\L}{\pi}\frac{E(k)}{k},
\ee
with $k^2=\lambda(\sqrt{q})$ the elliptic modulus 
and that, secondly, the complete elliptic integral of the second kind can be 
expressed as
\be
  \frac{E(k)}{k} = (1-k^2)\left(
  \frac{K(k)}{k} + \frac{\mathrm{d}K(k)}{\mathrm{d}k}
  \right)
\ee
in terms of the complete elliptic integral of the first kind. Then, the
result follows by simply plugging in 
$K(k) = \frac{1}{2}\pi\theta_3^2(\sqrt{q})$
and $k=\sqrt{\lambda(\sqrt{q})}=\theta_2^2(\sqrt{q})/\theta_3^2(\sqrt{q})$. 
With the help of the
appendix, the expression in Jacobi theta functions can then easily be
rewritten in terms of characters of the $c=-2$ triplet model.

It is worth noting that there is an interesting relation between the
modular form $c(\t)$ and the elliptic modulus $k$. To see this, let
us introduce modified characters which take into account the fermionic
nature of the $c=-2$ triplet model realized in terms of symplectic 
fermions. It turns out that the irreducible highest weight representations 
with conformal weights
$h=3/8$ and $h=1$ are spin doublets with all states having odd fermion number,
while the irreducible representations with weights $h=-1/8$ and $h=0$ are
spin singlets with all states having even fermion number. It is then
useful to introduce characters including Witten's index $F$, i.e.\
\be
  \chi_h(z,q) = \mathrm{tr}_{{\cal V}_h}z^Fq^{L_0-c/24}\,.
\ee
Defining in addition the quantity
\be
  \kappa(z,\t) = 
          \frac{\chi_{0}(z,q)+\chi_{1}(z,q)}
               {\chi_{-\frac{1}{8}}(z,q)+\chi_{\frac{3}{8}}(z,q)}\,,
\ee
we find the remarkable result
\be
  c(\t) = -\frac{\mathrm{i}\Lambda}{\pi}\frac{1}{\kappa(-1,\t)}\,,\ \ \ \
  k(\t) = 4\kappa^2(1,\t)\,.
\ee
It is worth mentioning that $\kappa(z,\t)$ is just the quotient of
the chiral partition functions of the twisted and untwisted sectors,
respectively, of the $c=-2$ ghost system \cite{Eholzer:1997se}.
Therefore, 
\be
  \begin{aligned}
  a(\t) &= \frac{\Lambda}{2\pi}
  \l 1-16\kappa^4(1,\t/2)\r \left( 
  \frac{ K(4\kappa^2(1,\t/2))}
  {\kappa^2(1,\t/2)}
        + \frac{1}{2\kappa(1,\t/2)}
        \frac{\mathrm{d}K(4\kappa^2(1,\t/2))}{\mathrm{d}\kappa(1,
	  \t/2)}
	\right)\,,\\
  a_D(\t) &= (\t-2)a(\t)+
  \frac{\mathrm{i}\Lambda}{\pi}\frac{1}
  {\kappa(-1,\t)}
        \,,
  \end{aligned}
\ee
Another guess is that the characteristic functions hint to topological
string theory, as there is a relation between Gromov-Witten invariants and
modular forms. This idea is strongly supported by \cite{Nekrasov:2003rj}. Only
recently there was published a paper by Huang and Klemm who indeed 
related the results
of \cite{Nekrasov:2003rj} to the topological B-model on a local CY 
\cite{Huang:2006si} and expressed the pure gauge $SU(2)$ prepotential up to
genus $6$ in terms of modular forms. The expressions for $a$ and $u$ calculated
there (which they do for the isogeneous curve) match ours but a factor 
of $2$ in the $q$-expansion. In another interesting
paper \cite{Aganagic:2006wq} Klemm and collaborators proved even more 
evidence for the suggestion above. 

\section*{Acknowledgements}
The work of MF is supported by by the European Union network
HPRN-CT-2002-00325 (EUCLID).


\appendix

\section{Useful Formulas and Definitions}\label{modu}

\begin{equation*}
  \begin{aligned}
    q&=\exp{(\i\pi\t)}\\
    \eta(q)&=q^{\frac{1}{12}}\prod_{n=1}^{\infty} (1-q^{2n}),\\
    \th_2(z,q)&=2q^{\4}\sum_{n=0}^{\infty}~q^{n(n+1)}\cos((2n+1)z),\\
    \th_3(z,q)&=1+2\sum_{n=1}^{\infty}~q^{n^2}\cos(2nz),\\
    \th_{\la,k}(q)&=\sum_{-\infty}^{\infty}q^{\frac{(2kn+\la)^2}{2k}},\\
  \end{aligned}
\end{equation*}
The characteristic functions in the $c=-2$ triplet model are the following:
\begin{equation*}
  \begin{aligned}
  \chi_{-\frac{1}{8}}(q)&=\frac{\th_{0,2}(q)}{\eta(q)},\\
  \chi_{\frac{3}{8}}(q)&=\frac{\th_{2,2}(q)}{\eta(q)},\\
  \chi_0(q)&=\frac{1}{2\eta(q)}(\th_{1,2}(q)-\eta(q)^3),\\
  \chi_1(q)&=\frac{1}{2\eta(q)}(\th_{1,2}(q)+\eta(q)^3),\\
  \chi_{\mc{R}_{0/1}}(q)&=2\eta(q)^{-1}\th_{1,2}(q),
  \end{aligned}
\end{equation*}
where the last character appears twice \cite{Gaberdiel:2001tr}, one for an
indecomposable
singlet highest weight representation $\mc{R}_0$ and the other for an
indecomposable doublet highest weight representation $\mc{R}_1$.
\par
\be
 \G^0(n):=\left\{ \begin{pmatrix} a&b\\c&d\end{pmatrix}~\in~SL(2,\mathbbm{Z})~
      :~b=0~\mathrm{mod}~n
      \right\}
\ee
The function $\la(\t)$ is the standard
elliptic modular function 
\be\label{ellmod}
  \la(\t):=\frac{e_3(\t)-e_2(\t)}{e_1(\t)-e_2(\t)}.
\ee
The $e_i$ are the branch points of the algebraic curve (\ref{weierstr}) 
and (\ref{J}) is invariant under $\G(2)$. $\la$ can be expressed in terms of
theta functions \cite{Bateman}
\be
  \la(\t)=\frac{\th_{2}^4(0,q)}{\th_{3}^4(0,q)}.
\ee
A nice property of the modular function is that its inverse can be given via
hypergeometric functions
\be\label{tauinhyp}
  \t=\i~\frac{{}_2F_1\l\2,\2;1;1-\la\r}{ {}_2F_1\l\2,\2;1;\la\r}.
\ee


\end{document}